%% file: main.tex
\documentclass[conference]{IEEEtran}
\usepackage{amsmath,amssymb,amsfonts}
\usepackage{algorithmic}
\usepackage{graphicx}
\usepackage{textcomp}
\usepackage{xcolor}
\usepackage{listings}
\usepackage{url}
\usepackage{cite}
\usepackage{color}
\usepackage{commath}
\usepackage{multirow}
\usepackage{balance}
\usepackage{cite}
\usepackage{comment}
\usepackage[shortlabels]{enumitem}

\def\BibTeX{{\rm B\kern-.05em{\sc i\kern-.025em b}\kern-.08em
    T\kern-.1667em\lower.7ex\hbox{E}\kern-.125emX}}
\makeatletter
\newcommand{\linebreakand}{%
  \end{@IEEEauthorhalign}
  \hfill\mbox{}\par
  \mbox{}\hfill\begin{@IEEEauthorhalign}
}

\newcommand{\inlinedComment}[2]
{\textcolor{#1}{\small\textbf{#2}}}

\begin{document}

\title{\huge SmartEmbed: A Tool for Clone and Bug Detection in Smart Contracts through Structural Code Embedding}

\author{\IEEEauthorblockN{Zhipeng Gao}
\IEEEauthorblockA{\textit{Monash University} \\
Melbourne, Australia \\
zhipeng.gao@monash.edu
}
\and
\IEEEauthorblockN{Vinoj Jayasundara}
\IEEEauthorblockA{\textit{Singapore Management University} \\
Singapore \\
vinojmh@smu.edu.sg
}
\and
\IEEEauthorblockN{Lingxiao Jiang}
\IEEEauthorblockA{\textit{Singapore Management University} \\
Singapore \\
lxjiang@smu.edu.sg
}
\hspace{-3cm}
\linebreakand
\IEEEauthorblockN{Xin Xia}
\IEEEauthorblockA{\textit{Monash University} \\
Melbourne, Australia \\
Xin.Xia@monash.edu
}
\and
\IEEEauthorblockN{David Lo}
\IEEEauthorblockA{\textit{Singapore Management University} \\
Singapore \\
davidlo@smu.edu.sg}
\and
\IEEEauthorblockN{John Grundy}
\IEEEauthorblockA{\textit{Monash University} \\
Melbourne, Australia \\
John.Grundy@monash.edu}
}

\maketitle

\begin{abstract}
\input{abstract}
\end{abstract}


\section{Introduction}
\label{sec:intro}
\input{intro}


\section{Approach}
\label{sec:approach}
\input{approach}

\section{Implementation Details \& Tool Usage}
\label{sec:implem}
\input{implem}

\section{Evaluation}
\label{sec:eval}
\input{eval}
\section{Summary and Future Work}
\label{sec:con}
\input{sum}

\section*{ACKNOWLEDGMENT}
\label{sec:ack}
\input{ack}

\balance
\bibliographystyle{unsrt}
\bibliography{main}

\end{document}

%% file: abstract.tex
Ethereum has become a widely used platform to enable secure, Blockchain-based financial and business transactions. However, a major concern in Ethereum is the security of its smart contracts. Many identified bugs and vulnerabilities in smart contracts not only present challenges to maintenance of blockchain, but also lead to serious financial loses.
There is a significant need to better assist developers in checking smart contracts and ensuring their reliability.
In this paper, we propose a web service tool, named {\sc SmartEmbed}, which can help Solidity developers to find repetitive contract code and clone-related bugs in smart contracts.
Our tool is based on code embeddings and similarity checking techniques. 
By comparing the similarities among the code embedding vectors for existing solidity code in the Ethereum blockchain and known bugs, we are able to efficiently identify code clones and clone-related bugs for any solidity code given by users, which can help to improve the users' confidence in the reliability of their code.
In addition to the uses by individual developers, {\sc SmartEmbed} can also be applied to studies of smart contracts in a large scale. When applied to more than 22K solidity contracts collected from the Ethereum blockchain, we found that the clone ratio of solidity code is close to 90\%, much higher than traditional software, and 194 clone-related bugs can be identified efficiently and accurately based on our small bug database with a precision of 96\%. 
{\sc SmartEmbed} can be accessed at \url{http://www.smartembed.net}. A demo video of {\sc SmartEmbed} is at 
\url{https://youtu.be/o9ylyOpYFq8}

%% file: intro.tex
In recent years, with the adoption and development of cryptocurrencies on distributed ledgers (a.k.a., blockchains), Ethereum has attracted increasingly attention as a blockchain platform. At the heart of the Ethereum platform are \emph{smart contracts}. 
A Smart contract is a computer program that can be triggered to execute any task when specifically predefined conditions are satisfied. A major concern in the Ethereum platform is the security of smart contracts. A smart contract in the blockchain often involves cryptocurrencies worthy of millions of dollars (e.g., DAO\footnote{\url{https://en.wikipedia.org/wiki/TheDAO(organization)}}, Parity\footnote{\url{https://paritytech.io/security-alert-2/}} and many more). Moreover, different from a traditional software program, the smart contract code is immutable after its deployment. They cannot be changed but may be killed when any security issue is identified within the smart contracts. This introduces challenges to blockchain maintenance and gives much incentive to hackers for discovering and exploiting potential problems in smart contracts, hence there is a very significant need to check and ensure the robustness of smart contracts before deployment.

Many prior works have investigated bug detection of smart contracts (e.g., \cite{luu2016making, tsankov2018securify, tikhomirov2018smartcheck}). A major disadvantage is that all these existing tools require certain bug patterns or specification rules defined by human experts. Considering the high stakes in smart contracts and race between attackers and defenders, it can be far too slow and costly to write new rules and construct new checkers in response to new bugs and exploits created by attackers.
Recently, there are also studies on clones and clone detection for Ethereum smart contracts (e.g., \cite{He2019,Eclone2018}). However, they use expensive symbolic transaction sketch or pair-wise comparisons which affect their efficiency and they are limited to clone detection.
Machine learning and deep learning techniques have been used for clone detection and bug detection problems (e.g. \cite{cclearner2017, Yang2015}) in traditional software programs too, but little has been applied for smart contracts.

In this paper, we present {\sc SmartEmbed}, a web service tool which can be accessible at \url{http://www.smartembed.net}. 
{\sc SmartEmbed} can efficiently and effectively check smart contracts for clones and bugs, and can evolve bug checking rules easily along with additions of new bugs.
The main idea of {\sc SmartEmbed} is two folds. (1) Code Embedding: utilizing basic program analyses and the availability of many open-source smart contracts, we encode each code element and bug pattern automatically, including their lexical, syntactical, and even some semantic information, into numerical vectors via techniques adapted from word embeddings (e.g., \cite{bojanowski2017enriching}).
(2) Similarity Checking: utilizing efficient similarity comparisons among the numerical vectors representing various kinds of code elements at various levels of granularity in smart contracts, we can detect clones similar to each other and bugs similar to known ones.

{\sc SmartEmbed} is unique in that it utilizes deep learning and similarity checking techniques to unify clone detection and bug detection together efficiently and accurately for Ethereum smart contracts.
When applied to more than 22K solidity contracts curated from the Ethereum blockchain, {\sc SmartEmbed} effectively tells us that the clone ratio of the Solidity code is at around 90\% and 194 out of 202 reported clone-related bugs are true bugs. 


\begin{figure*}\vspace{-0.0cm}
\centerline{\includegraphics[width=0.90\textwidth]{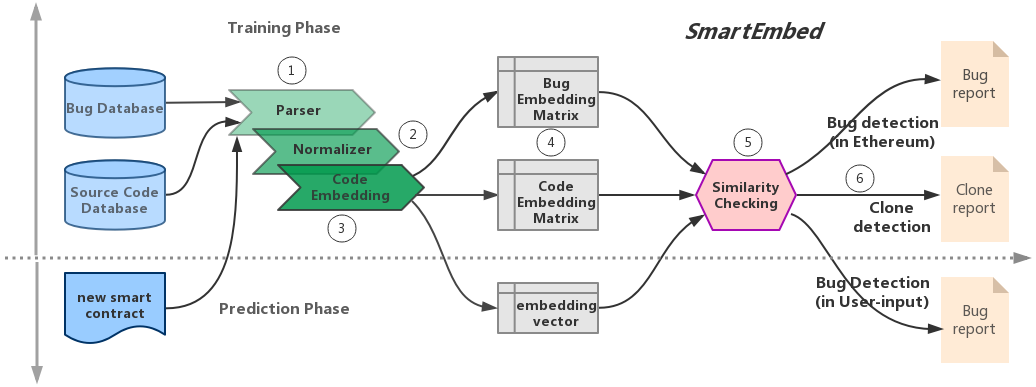}}
 \vspace*{-10pt}
\caption{Overview of Our Approach}
\label{fig:approach}
 \vspace{-0.0cm}
\end{figure*}

The rest of the paper is organized as follows.
Section~\ref{sec:approach} presents overall workflow of our approach and details of each step. 
Section~\ref{sec:implem} introduces the implementation details of our tool and its usages. 
Section~\ref{sec:eval} shows the experimental results of our evaluation.
Section~\ref{sec:con} concludes our work.

%% file: approach.tex
\definecolor{codegreen}{rgb}{0,0.6,0}
\definecolor{codegray}{rgb}{0.5,0.5,0.5}
\definecolor{codepurple}{rgb}{0.58,0,0.82}
\definecolor{backcolour}{rgb}{0.96,0.96,0.96}

\lstdefinestyle{mystyle}{
  backgroundcolor=\color{backcolour},   commentstyle=\color{codegreen},
  keywordstyle=\color{magenta},
  numberstyle=\tiny\color{codegray},
  stringstyle=\color{codepurple},
  basicstyle=\footnotesize,
  breakatwhitespace=false,         
  breaklines=true,                 
  captionpos=b,                    
  keepspaces=true,                 
  numbers=left,                    
  numbersep=5pt,                  
  showspaces=false,                
  showstringspaces=false,
  showtabs=false,                  
  tabsize=2
}
\lstset{style=mystyle}
\title{SmartEmbed}

\subsection{Overview}\label{AA}
Fig.\ref{fig:approach} illustrates the overall framework of {\sc SmartEmbed}. 
Based on code embeddings and similarity checking, 
{\sc SmartEmbed} targets two tasks in a unified approach: clone detection and bug detection. 
For clone detection, {\sc SmartEmbed} can identify similar smart contracts. For bug detection, based on our bug database, {\sc SmartEmbed} can detect bugs in the existing contracts in the Ethereum blockchain and/or in any smart contract given by solidity developers that are similar to any known bug in the database.
Our approach contains two phases: a model training phase and a prediction phase.

There are mainly 4 steps during the model training phase. We built a custom Solidity parser for smart contract source code. The parser generates an abstract syntax tree (ASTs) for each smart contract in our collected dataset, and serializes the parse tree into a stream of tokens depending on the types of the tree nodes (step 1). After that, the normalizer reassembles the token streams to eliminate nonessential differences (e.g., the stop word, values of constants or literals) between smart contracts (step 2). The output token streams are then fed into our code embedding learning module, and each code fragment is embedded into a fixed-dimension numerical vector (step 3). After the code embedding learning step, all the source code is encoded into the source code embedding matrix; in the meanwhile, all the bug statements we collected are encoded into the bug embedding matrix (step 4).

In the prediction phase, any given new smart contract is turned into embedding vectors by going through the steps 1,2,3 and utilizing the learned embedding matrices. Similarity comparison is performed between the embeddings for the given contract and those in our collected database (step 5), and similarity thresholds are used to govern whether a code fragment in the given contract will be considered as code clones or clone-related bugs (step 5-6).

\subsection{Details}
\subsubsection{Parsing}
{\sc SmartEmbed} employs ANTLR\footnote{\url{https://www.antlr.org/}} and a custom Solidity grammar\footnote{\url{https://github.com/solidityj/solidity-antlr4}} to generate ASTs for each smart contract. 
Listing 1 shows a simple example of a smart contract defined in Solidity.
Depending on the types of the tree nodes, the ASTs is serialized differently for contract-level and statement-level program elements to capture structural information in and around the focal elements. 

\begin{lstlisting}[language=Java, caption=An Example Solidity Program]
pragma solidity ^0.4.15;

contract Overflow {
    uint private r=0;
    
    function addValue(uint value) returns (bool){
        // possible overflow
        r += value; 
    } 
}
\end{lstlisting}

\vspace{0.1cm}\noindent {\em  \bf Contract Level Parsing:} All terminal tokens are extracted from the ASTs by an in-order traversal. The contract level parsing result of the sample code is shown below.
\begin{lstlisting}
1_10 : pragma solidity ^ versionliteral ; contract Overflow { uint private r = 0 ; function addValue ( uint value ) returns ( bool ) {  r  += value ; } }
\end{lstlisting}
\vspace{0.1cm}\noindent {\em  \bf Statement Level Parsing:} For statement parsing, more structural information (containment and neighbouring) as well as some semantic information (data-flow) is added to the sequences. The statement level parsing result of line 8 is given as follows. 
\begin{lstlisting}
 8_8 : sourceUnit contractDefinition contractPart functionDefinition block statement simpleStatement r += value ; function addValue add value ( uint value ) returns ( bool ) contract Overflow overflow { }
\end{lstlisting}

\subsubsection{Normalization} {\sc SmartEmbed} normalizes the parsing sequence to remove some semantic-irrelevant information. All simple variables, non-essential punctuation marks and different type of constants are replaced or removed. 
The following code snippet exemplifies the operation of this step:
\begin{lstlisting}
    uint private r = 0 ; 
     ==> 
    uint private SimpeVar = decimalnumber 
\end{lstlisting}

\subsubsection{Code Embedding Learning} {\sc SmartEmbed} embeds code elements and bug patterns, including their lexical, syntactical, and some semantic information into numerical vectors via techniques adapted from word embeddings. We choose Fasttext\cite{bojanowski2017enriching} as the code embedding algorithm as it performed on par or better compared with traditional word2vec.

\vspace{0.1cm}\noindent {\itshape\bfseries Token Embedding:} The normalized token streams with structural information generated by the normalizer for the solidity contracts 
are used as the training corpus.
We adapted the Fasttext algorithm to train code embedding models.
After the training, each token in the training corpus, including the tokens representing structural information, is mapped to fixed-dimension vector with real values. 

\vspace{0.1cm}\noindent {\itshape\bfseries Higher Level Embedding:} Based on the basic vector representation for each token, the code embeddings for higher-level code fragments (e.g., statements, functions, subcontracts, and contracts) are composed together. To be more specific, we define the code embeddings for a particular code fragment as the summation of the embeddings of all its constituent tokens.

\subsubsection{Embedding Matrices}
By stacking individual vectors together, we obtain a source code embedding matrix $\mathbf{C^{c \times d}}$ for clone detection and a bug statement embedding matrix $\mathbf{B^{b \times d}}$ for bug detection.

\vspace{0.1cm}
\noindent {\itshape\bfseries Source Code Embedding Matrix} $\mathbf{C^{c \times d}}$: The first dimension $c$ is the total number of contracts; the second dimension $d$ is the code embedding size we set previously.
The $i$$^{th}$ element $\mathbf{C_i}$ ($i = 1, 2, ..., c$) is the vector representation for the $i$$^{th}$ contract.

\noindent {\itshape\bfseries Bug Statement Embedding Matrix} $\mathbf{B^{b \times d}}$: The first dimension $b$ corresponds to the total number of bug statements in our bug database, and each row of the matrix, i.e., $\mathbf{B_i}$ ($i = 1, 2, ..., b$)  represents the code embedding for a specific bug statement.

\subsubsection{Similarity Checking}
\label{sec:similarity}
We define a similarity metric, which is used in the downstream tasks of clone detection and bug detection.

\vspace{0.1cm}
\noindent\textbf{\textit{Definition}}: Let $C_1$ and $C_2$ be two code fragments, and $e_1$ and $e_2$ be their corresponding code embeddings.
We define the semantic distance as well as similarity between the two code snippets as follows:
\begin{equation}
Distance(C_1, C_2)= \frac{Euclidean(e_1, e_2)}{\norm{e_1 }+ \norm{e_2}} \label{eq2}
\end{equation}
\begin{equation}
Similarity(C_1, C_2)= 1 - Distance(C_1, C_2) \label{eq3}
\end{equation}
Given any two code fragments $C_i$ and $C_j$, if their similarity score is over a specific similarity threshold $\delta$, $C_i$ and $C_j$ are viewed as a clone pair. 

\subsubsection{Clone Detection and Bug Detection}
Both clone detection and bug detection tasks can be viewed as variants of the problem of finding ``similar'' code, depending on the definition of similarity. For clone detection, we measure the similarity between pairs of smart contracts, and identify them as clones if the similarity score is over a predefined threshold for clones.
For bug detection, we search for code fragments in given contracts that are more similar to the known bugs than a predefined threshold for bugs.

%% file: implem.tex
We have implemented {\sc SmartEmbed} as a standalone web service to facilitate Solidity developers in checking their smart contracts. The source code and data can be found in our Github repository\footnote{\url{https://github.com/beyondacm/SmartEmbed}}.

\smallskip
\noindent{\itshape\bfseries Data Collection}.
We collected 22,275 verified Solidity smart contracts using EtherScan\footnote{https://etherscan.io/}, which is a block explorer and analytics platform for Ethereum.
The contracts contain 135,239 subcontracts, 631,261 functions, around 2 million statements, and more than 7 million lines of code.
In the meanwhile, we collected 22 well-known vulnerable smart contracts and pinpointed 37 buggy statements in the contracts, which served as the bug database for {\sc SmartEmbed}.

\smallskip
\noindent{\itshape\bfseries Backend Model}.
The collected source code of contracts are inputted into the workflow of our approach described in Section ~\ref{sec:approach}, and the output is the code embeddings which are used as the backend model for similarity checking.

\begin{figure}[t]
    \centerline{\includegraphics[width=0.43\textwidth]{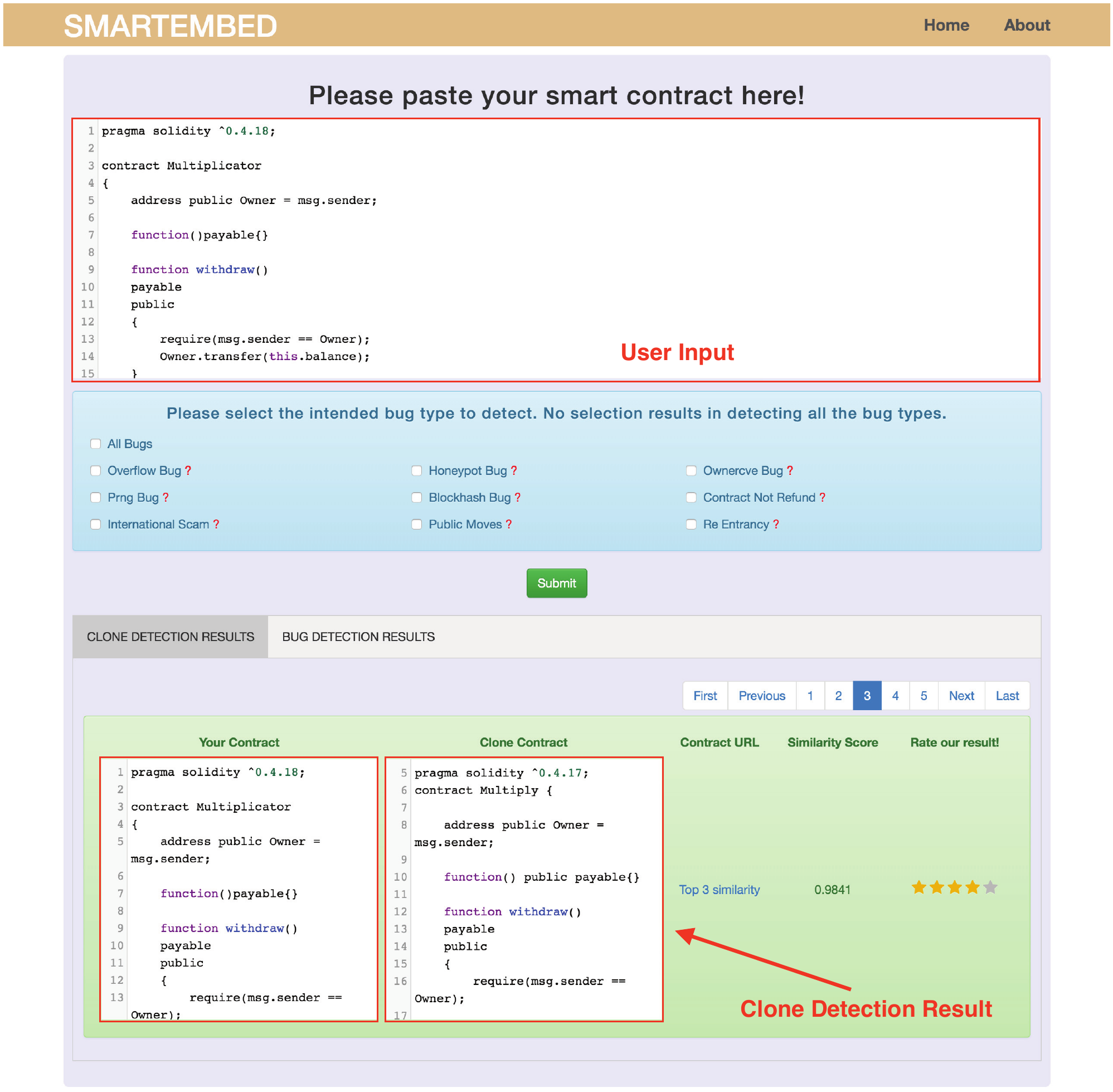}}
\vspace*{-10pt}
\caption{Sample Results of Clone Detection}\vspace*{-0.2cm}
\label{fig:tool-clone}
\end{figure}

\begin{figure}[t]
    \centerline{\includegraphics[width=0.43\textwidth]{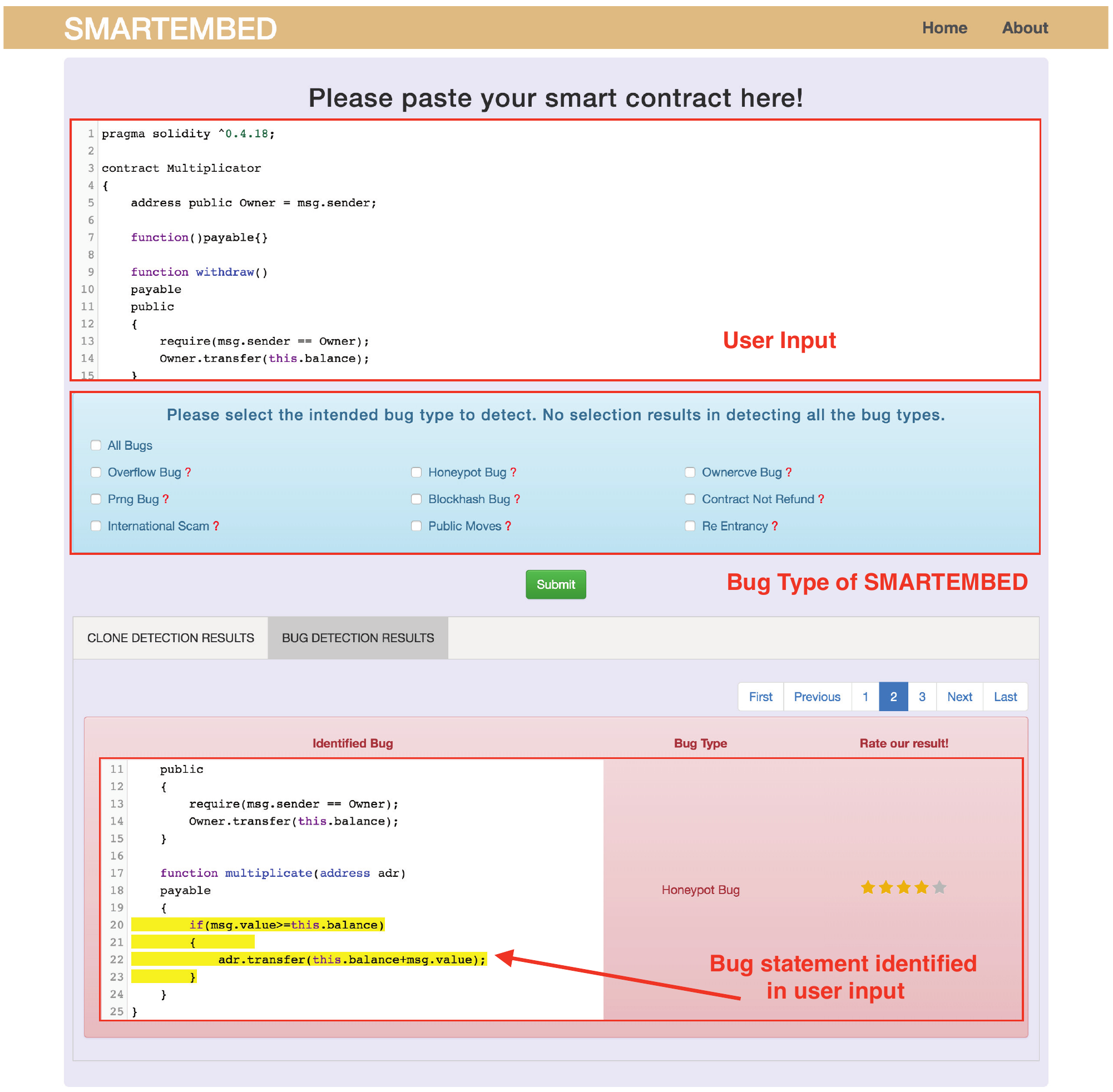}}
\vspace*{-10pt}
\caption{Sample Results of Bug Detection}\vspace*{-0.2cm}
\label{fig:tool-bug}
\end{figure}

\smallskip
\noindent{\itshape\bfseries Frontend User Interface}.
On the user interface, {\sc SmartEmbed} provides an input box for Solidity developers to submit their source code (cf.\ Fig.~\ref{fig:tool-clone} and Fig.~\ref{fig:tool-bug}). 
After a Solidity developer submits his/her source code to the server, the source code is parsed and normalized, then the contract and each statement is converted into a vector by our backend model for similarity checking.
The outputs are organised into two separate result tabs. 
For the clone detection result tab, {\sc SmartEmbed} returns top-5 most similar clone contracts in our code base together with the similarity scores and links to their locations in EtherScan (cf.\ Fig.~\ref{fig:tool-clone}).
For the bug detection result tab, {\sc SmartEmbed} highlights the buggy lines in the submitted source code and reports the bug types back to the developer (cf.\ Fig.~\ref{fig:tool-bug}).

%% file: eval.tex
We compared {\sc SmartEmbed} with two well-known tools that are specific for clone detection (DECKARD~\cite{jiang2007deckard} extended for Solidity) and bug detection (SmartCheck~\cite{tikhomirov2018smartcheck}) respectively.
For clone detection, we run DECKARD and {\sc SmartEmbed} against 22,725 smart contracts source code, 
the experiment results show that both tools identified around 6.6 million lines of code as code clones, while the total lines are just 7.3 million, which means the clone ratio of solidity code is at around 90\%, much higher than traditional software. One main reason for introducing clones is the irreversibility of smart contracts stored in Ethereum blockchain.
By manually checking some clones detecked by our approach but not by DECKARD, 
we found code clones such as type-III or even type-IV semantic code clones can also be detected, which means {\sc SmartEmbed} is highly effective to identify the code clones in smart contract. 
For bug detection, {\sc SmartEmbed} can identify clone related bugs in Ethereum blockchain more efficiently and accurately.
When the similarity threshold is set to 0.95, our tool reports 202 clone related bugs, we manually validate these candidate bugs and 194 of which are labelled as true bugs, while SmartCheck can only detect 117 of these verified bugs by using the same bug pattern type within our bug database. 

%% file: sum.tex
This paper presented {\sc SmartEmbed}, a web service tool for detecting code clones and bugs in smart contracts accurately and efficiently.
It develops a code embedding technique for tokens and syntactical structures in Solidity code and utilizes similarity checking to search for ``similar'' code satisfying certain thresholds.
The approach is automated on the contract and bug data collected from the Ethereum blockchain.
It helps developers to find repetitive contract code and clone-related bugs in existing contracts. 
It also helps to efficiently validate given smart contracts against known set of bugs without the need of manually defined bug patterns.
Its backend model can be easily updated to recognize new contract clones and new kinds of bugs when the contract code and bugs evolve.
In the future, we plan to enrich the contract and bug databases so that {\sc SmartEmbed} can detect more clones and bugs.



%% file: ack.tex
This research is supported by the Singapore Ministry of Education (MOE) Academic Research Fund (AcRF) Tier 1 grant from SIS at SMU. We also thank the anonymous reviewers for their insightful comments and suggestions.